# Microtargeted propaganda by foreign actors: An interdisciplinary exploration

Ronan Ó Fathaigh *, Tom Dobber**,
Frederik Zuiderveen Borgesius***,
and James Shires****

## Abstract

This article discusses a problem that has received scant attention in literature: microtargeted propaganda by foreign actors. Microtargeting involves collecting information about people, and using that information to show them targeted political advertisements. Such microtargeting enables advertisers to target ads to specific groups of people, for instance people who visit certain websites, forums, or Facebook groups. This article focuses on one type of microtargeting: microtargeting by foreign actors. For example, Russia has targeted certain groups in the US with ads, aiming to sow discord. Foreign actors could also try to influence European elections, for instance by advertising in favour of a certain political party. Foreign propaganda possibilities existed before microtargeting. This article explores two questions. In what ways, if any, is microtargeted propaganda by foreign actors different from other foreign propaganda? What could lawmakers in Europe do to mitigate the risks of microtargeted propaganda?

*Institute for Information Law, University of Amsterdam, Amsterdam, the Netherlands
**Amsterdam School of Communication Research, University of Amsterdam, Amsterdam, the Netherlands
***Interdisciplinary Hub for Security, Privacy, and Data Governance (iHub) and Institute for Computing and Information Sciences (iCIS), Radboud University Nijmegen, Nijmegen, the Netherlands
****Institute for Security and Global Affairs, University of Leiden, Leiden, the Netherlands

**Corresponding author:**
Ronan Ó Fathaigh, Institute for Information Law, Faculty of Law, University of Amsterdam, Nieuwe Achtergracht 166, 1018 WV Amsterdam, the Netherlands.
Email: r.f.fahy@uva.nl

## 1. Introduction

As technology evolves, foreign adversaries' means to intervene in democratic processes evolve as well.[1] For instance, foreign actors interfered in the presidential election in the US in 2016, including by buying political ads on social media in the names of US persons and entities.[2] Indeed, according to the European Commission, there were attempts at foreign interference in the 2019 European Parliament elections.[3] Techniques like political microtargeting, which involves collecting information about people, and using that information to show them tailored political advertisements, can be used to spread propaganda. Microtargeting enables advertisers to target ads to specific groups of people. Advances in artificial intelligence enable foreign actors to create potent propaganda, such as AI-generated fake videos: deepfakes. By combining microtargeting techniques with such new forms of propaganda, foreign actors could send the 'right' propaganda message to the 'right' person and influence the integrity of fundamental democratic processes, including European elections.[4]

Extant research has focused mainly on the role of microtargeting within democratic states, focusing on scenarios in which microtargeting will be used by legitimate domestic political actors, such as domestic political parties and candidates.[5] But while these domestic political actors mostly play by the rules, foreign actors have less reason to conform to domestic norms and regulation. Of course, foreign propaganda aimed at influencing democratic processes existed before microtargeting. However, microtargeting brings new dimensions to the problem. This article thus focuses on the following questions. In what ways, if any, is microtargeted propaganda by foreign actors different from other forms of propaganda? What could lawmakers in Europe do to mitigate the risks of microtargeted propaganda by foreign actors?

For brevity, we also speak of 'foreign microtargeting' when referring to 'microtargeted propaganda by foreign actors'. We combine insights from different disciplines: law, political science, international relations and communication science. The article does not discuss all aspects of microtargeting. Outside the scope are, for instance, general questions relating to the promises and threats of microtargeting for democracy.[6]

The article is structured as follows. Section 2 discuss the notion of propaganda, and how international law defines propaganda. Section 3 discusses to what extent foreign microtargeting differs from other forms of propaganda, and gives examples of how foreign actors have used, and could use, microtargeting. Section 4 explores what lawmakers in the EU could do to mitigate risks of foreign microtargeted propaganda. Section 5 highlights complications when regulating such practices. Section 6 concludes.

---

1. By 'foreign' we mean outside the political community (usually a state) where democratic processes take place – see discussion at the end of section 2 for more details. To limit the scope of the analysis, we focus only on democratic processes, not other political systems. Foreign intervention in non- or semi-democratic political processes using the means discussed here is highly plausible but raises very different questions of fundamental rights and international law.
2. US Department of Justice, *Report on the Investigation into Russian Interference in the 2016 Presidential Election, Special Counsel Robert S. Mueller, III* (US Department of Justice, 2019), https://www.justice.gov/storage/report.pdf.
3. European Commission, Communication on the European Democracy Action Plan, COM/2020/790 final, p. 3. See also European Commission and the High Representative, Joint Communication on the Report on the implementation of the action plan against disinformation, JOIN(2019) 12 final, p. 3.
4. J.-B. Jeangène Vilmer et al., *Information Manipulation: A Challenge for Our Democracies* (French Ministry for Europe and Foreign Affairs and the Ministry for the Armed Forces, 2018); and European Commission, Tackling online disinformation: a European Approach, COM(2018) 236 final.
5. F.J. Zuiderveen Borgesius et al., 'Online Political Microtargeting: Promises and Threats for Democracy', 14 *Utrecht Law Review* (2018), p. 82–96.
6. See on such topics: ibid., p. 82.

## 2. Defining propaganda

Propaganda is a slippery term with a long history, often used in common language simply to denote bias. It is now morally charged, with clear pejorative connotations, although this was not always the case – in World War II, propaganda was an accepted part of the war effort in the US and UK.[7] Despite growing negative Cold War associations, theorists writing in the second half of the twentieth century still defined propaganda, without value judgement, as communication 'that invites us to respond emotionally, emphatically, more or less immediately, and in an either-or manner'.[8] More widely, religious and other ideological movements have long used the term neutrally or even positively: as Jowett and O'Donnell indicate in their historical review of the concept, pointing to Christian thought from the Crusades to the Reformation, 'religious ideology can also propagandize for the good of society'.[9] Indeed, the social and political, rather than biological ('to propagate') use of the Latin term comes from a commission established by Pope Gregory XIII in the late 16th century.[10] Many scholars seek to accommodate these normative and conceptual diversions by distinguishing between types or 'shades' of propaganda.[11] We use the term because it captures the essence of the problem of illegitimate public information manipulation,[12] and usefully connects modern digital issues to their longer context.[13] To provide some clarity among the broad variety of connotations attached to the term over centuries of use, and to make our discussion most relevant for readers of this journal, we focus on narrower definitions of propaganda in international law and human rights.

To understand the notion of propaganda under the international human rights system, we can turn to the UN Special Rapporteur on freedom of expression, who has warned that disinformation and propaganda are 'global problems', which are now 'exacerbated in the digital age'.[14] The Special Rapporteur considers that current forms of propaganda and disinformation aim to undermine the public's trust in information, the media, and public institutions. While some disinformation may have non-governmental origins, much of the current disinformation and propaganda has governmental (or quasi-governmental) origins, with current digital tools enabling disinformation and propaganda to 'spread quickly, easily'.[15]

---

7. For historical treatments and overviews of the development of propaganda, see e.g. E. Bernays, *Propaganda* (Ig. Publishing, 2005) (originally published 1928); and P. Baines et al., (eds.), *The SAGE Handbook of Propaganda* (SAGE Publications Ltd, 2019). In the 1940s, Bernays himself was hired by the United Fruit Company to mount a propaganda campaign as part of a coup in Guatemala, with his views of effectiveness of certain forms of propaganda influencing later CIA thinking on this subject. Ronald J. Deibert, *Reset: Reclaiming the Internet for Civil Society* (House of Anansi Press Ltd, 2020), p. 116.
8. Renee Hobbs, '"A Most Mischievous Word": Neil Postman's Approach to Propaganda Education", 2 *Harvard Kennedy School Misinformation Review* (2021).
9. Garth S. Jowett and Victoria O'Donnell, *Propaganda & Persuasion* (7th edition, SAGE Publications Ltd, 2018), p. 59.
10. Erwin W. Fellows, '"Propaganda:" History of a Word', 34(3) *American Speech* (1959), p. 182.
11. D. Omand, 'The Threats from Modern Digital Subversion and Sedition', 3(1) *Journal of Cyber Policy* (2018), p. 5; C. Walton, 'Spies, Election Meddling, and Disinformation: Past and Present', 26 *Brown Journal of World Affairs* (2019), p. 107.
12. D. Susser et al., 'Online Manipulation: Hidden Influences in a Digital World', 4 *Georgetown Technology Law Review* (2019), p. 1.
13. P. Pomerantsev, *This Is Not Propaganda: Adventures in the War Against Reality* (Faber & Faber, 2019).
14. Special Rapporteur on the Promotion and Protection of the Right to Freedom of Opinion and Expression, Statement to the General Assembly – Third Committee, 24 October 2017, https://www.ohchr.org/en/NewsEvents/Pages/DisplayNews.aspx?NewsID=22300&LangID=E.

Indeed, current concerns over propaganda led the four international special mandates on freedom of expression from the UN, the Organization for Security and Co-operation in Europe (OSCE), the Organization of American States (OAS), and the African Commission on Human and Peoples' Rights, to issue a Joint Declaration on 'fake news', disinformation and propaganda.[16] The Joint Declaration contained an authoritative description of disinformation and propaganda, saying that state actors should not make, sponsor, encourage or further disseminate statements 'which they know or reasonably should know to be false (disinformation)' or 'which demonstrate a reckless disregard for verifiable information (propaganda)'.[17] The Joint Declaration explains:

> [D]isinformation and propaganda are often designed and implemented so as to mislead a population, as well as to interfere with the public's right to know and the right of individuals to seek and receive, as well as to impart, information and ideas of all kinds, regardless of frontiers, protected under international legal guarantees of the rights to freedom of expression and to hold opinions. (…)
>
> [S]ome forms of disinformation and propaganda may harm individual reputations and privacy, or incite to violence, discrimination or hostility against identifiable groups in society.[18]

Scholars who have examined various definitions of propaganda, such as Bayer et al., consider this a clearer treatment than many scholarly definitions,[19] with its focus on the 'intent' of the propaganda, rather than its content, and thus excluding from its scope many 'legitimate forms of persuasion'.[20] Propaganda is illegitimate, in this definition, because it misleads, thereby fitting within Susser, Roessler and Nissenbaum's broader category of 'online manipulation' as 'hidden influence – the covert subversion of another person's decision-making power'.[21] To anticipate the argument below, microtargeted propaganda is concerning precisely because it deploys techniques that make what Susser et al. call 'covert subversion' more effective and efficient.[22]

---

15. Ibid.
16. UN Special Rapporteur on Freedom of Opinion and Expression, the Organization for Security and Co-operation in Europe (OSCE) Representative on Freedom of the Media, the Organization of American States (OAS) Special Rapporteur on Freedom of Expression and the African Commission on Human and Peoples' Rights (ACHPR) Special Rapporteur on Freedom of Expression and Access to Information, Joint Declaration on freedom of expression and 'fake news', disinformation and propaganda, FOM.GAL/3/17, 3 March 2017, preamble, https://www.osce.org/fom/302796?download=true ('Joint Declaration').
17. Ibid.
18. Ibid., section 2(c).
19. For an overview, see J.M. Grygiel, 'Algorithmic Propaganda: How Facebook Meddles with Democracy', 25 *Communications Law* (2020), p. 23.
20. J. Bayer et al., *Disinformation and Propaganda – Impact on the Functioning of the Rule of Law in the EU and its Member States* (European Parliament, 2019), p. 27.
21. D. Susser et al., 'Online Manipulation: Hidden Influences in a Digital World', 4 *Georgetown Technology Law Review* (2019), p. 3. Susser et al., do not use the term 'propaganda' in their article. For a contrary view emphasizing the 'obvious' nature of propaganda, see Tim Wood, 'Propaganda, Obviously: How Propaganda Analysis Fixates on the Hidden and Misses the Conspicuous', 2(2) *Harvard Kennedy School Misinformation Review* (2021).
22. For a useful overview of different mechanisms of coercive/persuasive and conscious/subconscious influence, see B.M.J. Pijpers and P.A.L. Ducheine, 'Influence Operations in Cyberspace – How They Really Work', *The Hague: The Hague Program for Cyber Norms* (2020).

The following elements can be deduced from the above description of propaganda: (i) disseminating information (ii) which is designed to (iii) mislead a population, (iv) interfere with the public's right to know and the right of individuals to seek and receive, as well as to impart, information and ideas of all kinds; and (v) undermine the public's trust in information, the media, or public institutions.

However, while disinformation and propaganda are a major concern under the international human rights system, and particularly for the public's right to receive information, international human rights law only prohibits certain forms of propaganda. As the OSCE Representative on Freedom of the Media has stated, given the 'broadness and vagueness' of the term propaganda, and its 'direct link' to political freedom of expression, a blanket prohibition on propaganda would violate international human rights standards on the protection of freedom of expression.[23]

Only two types of propaganda are explicitly prohibited under international human rights law: (a) propaganda for war, and (b) advocacy of national, racial or religious hatred. Indeed, the International Covenant on Civil and Political Rights places an obligation on States to prohibit by law '[a]ny propaganda for war' and '[a]ny advocacy of national, racial or religious hatred that constitutes incitement to discrimination, hostility or violence'.[24] Similarly, under the International Convention on the Elimination of All Forms of Racial Discrimination, States must prohibit all propaganda activities that 'promote and incite racial discrimination'.[25] As such, most countries throughout the world, and all European countries have national legislation prohibiting incitement to discrimination, hatred and violence.[26] Indeed, in the European Union, the 2008 Council Framework Decision on combating certain forms and expressions of racism and xenophobia by means of criminal law requires EU member states to prohibit 'publicly inciting to violence or hatred directed against a group of persons or a member of such a group defined by reference to race, colour, religion, descent or national or ethnic origin'.[27]

The UN Human Rights Committee elaborated upon the definition of prohibited propaganda as: all forms of propaganda 'threatening or resulting in an act of aggression or breach of the peace contrary to the Charter of the United Nations', and 'any advocacy of national, racial or religious hatred that constitutes incitement to discrimination, hostility or violence, whether such propaganda or advocacy has aims which are internal or external to the State concerned'.[28]

Such traditional propaganda was prevalent in the Ukraine crisis in 2014, during which the OSCE Representative on Freedom of the Media issued a Communiqué on propaganda in time of conflict.[29]

---

23. OSCE Representative on Freedom of the Media, Communiqué on propaganda in times of conflict, 15 April 2014, https://www.osce.org/fom/117701.
24. International Covenant on Civil and Political Rights, 19 December 1966, 999 UNTS 171, Article 20(1).
25. International Convention on the Elimination of All Forms of Racial Discrimination, 4 January 1965, 660 UNTS 195, Article 4(b).
26. For example, Article 261 bis of the Swiss Criminal Code ('Any person who publicly stirs up hatred or discrimination against a person or group of persons on the grounds of their race, ethnic origin or religion … shall be punishable by a custodial sentence of up to three years or a fine'). For a discussion, see ECtHR, *Perinçek v. Switzerland*, Grand Chamber Judgment of 15 October 2015, Application No. 275510/08.
27. Council Framework Decision 2008/913/JHA of 28 November 2008 on combating certain forms and expressions of racism and xenophobia by means of criminal law, Article 1.
28. UN Human Rights Committee (HRC), *CCPR General Comment No. 11: Article 20 Prohibition of Propaganda for War and Inciting National, Racial or Religious Hatred*, 29 July 1983, section 2.
29. OSCE Representative on Freedom of the Media, Communiqué on propaganda in times of conflict, 15 April 2014, https://www.osce.org/fom/117701.

The Representative emphasised that propaganda, coupled with a deterioration of media freedom, often go together to 'fuel a conflict', and can 'contribute to its escalation', noting that '[t]oday in the 21st century, as it was in the past, state media is the main vehicle of propaganda'.[30]

So far, this section considered propaganda by both domestic and foreign actors equally, and scholarship on propaganda more broadly (such as the UN Human Rights Committee definition above on 'internal' and 'external' actions). However, in this article we focus only on propaganda by foreign actors. By 'foreign' we mean outside the political community (usually a state) where democratic processes take place. By using this term, we do not mean to imply it is easy to distinguish foreign and domestic influences in democratic processes; such an exercise is difficult in an era of thick transnational connections between state and local governments, corporations, and other non-governmental institutions. The political community, moreover, does not have to be a state – the EU would also meet this definition, and so it is possible to have 'foreign' influence in European elections (we discuss difficult edge cases of such influence in section 4.B below). Our use of 'foreign' requires only that some actors be illegitimate participants in democratic processes due to their exclusion from the relevant political community.

An important grey area in this definition arises at the sub-state, rather than supra-state, level. States themselves are composed of many – often contradictory – political communities, amalgamated through a combination of common interest and the more violent exertion of power, both historical and contemporary. Consequently, the legitimacy of political communication by a dominant central government in elections at local or other sub-state levels is frequently contested. Such contests appear most starkly in elections or referenda on devolution and secession, such as in Northern Ireland, Scotland or Catalonia.[31] In such cases, central government campaigns (from London or Madrid) may well be seen as virtually 'foreign' propaganda. However, this is not the norm, and most local government elections in democracies are connected explicitly to national party politics, signalling membership of the same political community at a state level.[32]

A different aspect of this problem emerges in relation to global movements like communism, or less coordinated affinities such as those between 'green' parties focusing on climate and environmental issues worldwide. On one hand, the Communist International was widely – and as Rid demonstrates, sometimes wrongly – perceived as no more than a front for Russian propaganda.[33] In contrast, green parties have rarely attracted such criticism, and so coordinated campaigns between green parties in different states may not be seen as foreign propaganda to the same extent. The legitimacy of this complex relationship between global movements and local organizations can only be decided in specific contexts, based on the location, direction and motivation of both movement leaders and domestic partners.[34]

In this way, the distinction between foreign and domestic propaganda motivating this article foregrounds difficult questions of political community and legitimacy that are often simplified in media discourses around the dangers of online political influence. To be clear, we limit the scope of this article to electoral contexts, and so do not address the legitimacy of propaganda by foreign actors outside such contexts; for example, the controversial PR campaigns by the Saudi Arabian Crown Prince Muhammad bin Salman surrounding his tours of the US and UK in 2018.[35] Whatever the appropriate boundaries of intervention in domestic political discourse in general, it is an accepted diplomatic norm that these boundaries are stricter around elections.[36] Of course, the right to participate in elections is itself a form of political boundary-drawing and so hotly contested, especially with regard to immigration and citizenship.

---

30. Ibid.
31. See e.g. D. Turp et al., *The Catalan Independence Referendum: An Assessment of the Process of Self-Determination* (IRAI, 2017). L. Whitehead, 'The Hard Truths of Brexit', 31(2) *Journal of Democracy* (2020).
32. For an influential dissection of the legitimacy of various 'levels' of government in domestic politics, see P. Hirst, *From Statism To Pluralism* (Routledge, 1997).
33. T. Rid, *Active Measures: The Secret History of Disinformation and Political Warfare* (Profile Books, 2020).
34. For a classic discussion, see Margaret E. Keck and Kathryn Sikkink, *Activists Beyond Borders: Advocacy Networks in International Politics* (Cornell University Press, 1998).

Given the inevitably partial nature of this brief discussion on foreignness and the nature of political community, we conclude this section by reiterating why foreign actors are the focus of this article. First, as stated in the introduction, foreign actors have less reason to conform to domestic norms and regulation. In fact, insofar as it does not limit the effectiveness of their propaganda, foreign actors have incentives *not* to conform, to gain comparative advantage against those with greater local knowledge.[37] Second, whether or not foreign propaganda is always more damaging than domestic versions – a question difficult to settle in the abstract – it presents a far more challenging regulatory and legal problem due to difficulties in jurisdiction and enforcement (to which we return below), as well as raising fundamental concerns of sovereignty and self-determination.[38] For these reasons, foreign propaganda is high on the agenda of national governments and at a European level, following high-profile instances of election interference in Europe and the US attributed to other states. We specify in the following sections where the discussion applies equally to foreign and domestically produced propaganda, and where the two diverge, especially in terms of potential actions to mitigate the risks.

## 3. Microtargeted propaganda: what is new?

Political microtargeting (microtargeting for short)[39] involves collecting information about people, and using that information to show them targeted political advertisements. Propaganda without microtargeting is relatively static. Multiple citizens are exposed to a deceitful advertisement, a false news story or a photoshopped image. By itself, these types of static propaganda are problematic as they can cause citizens to think or act on the basis of lies, deceit or manipulation.[40] Microtargeting can amplify these negative effects of propaganda.

---

35. See especially the decision by the UK media regulator to ban a concurrent TV commercial due to its classification as 'political advertising'. A similar online PR campaign was not within the scope of this decision. Ofcom, *Broadcast and On Demand Bulletin* (Ofcom [UK media regulator], 2018), p. 9. More broadly, a key function of diplomatic missions is to represent a foreign state's perspective in domestic discussions, both public and private.
36. An indication of the strength of this norm comes indirectly through responses – domestic and international – to its flouting by former US President Donald Trump. See e.g. Jonathan Allen, 'Trump Normalizes Foreign Election Intervention Ahead of 2020', *NBC News*, 3 June 2019. www.nbcnews.com/politics/white-house/trump-normalizes-foreign-election-intervention-ahead-2020-n1013091.
37. For evidence of this approach, see the detailed interviews in Mona Elswah and Philip N Howard, '"Anything That Causes Chaos": The Organizational Behavior of Russia Today (RT)', 70(5) *Journal of Communication* (2020), p. 623.
38. Nicholas Tsagourias, 'Electoral Cyber Interference, Self-Determination and the Principle of Non-Intervention in Cyberspace', in Dennis Broeders and Bibi van den Berg (eds.), *Governing Cyberspace: Power, Behavior, and Diplomacy* (Rowman & Littlefield Publishers, Inc., 2020), p. 45.
39. The distinction between political and non-political advertising is difficult to draw cleanly for platforms such as Facebook, because they define it as referring to a 'political candidate or issue'. Of course, what is a political issue changes depending on context, and the boundaries of politics are themselves a political question. We seek only to ensure this article is not focused on 'purely' commercial advertising, rather than limiting 'political' advertising to elections or similar key events.

Microtargeted propaganda has several new aspects, compared to traditional propaganda methods. We categorise the new aspects of microtargeted propaganda in three groups: effectiveness, efficiency and opacity. In short, microtargeting techniques help expose the right person to the right message, at least from the perspective of the sender, while hiding the origin of the message. We then explore how microtargeted propaganda intersects with other relevant technological advances, especially deepfakes.

Although we focus only on foreign actors in the following sections, internal political actors can also use political microtargeting to achieve their aim, and so the following novel characteristics of microtargeted propaganda are also applicable to those actors. Even though microtargeted propaganda by domestic actors is outside the scope of this article, some of the regulatory solutions we suggest in Section 4 could alleviate the problems posed by both domestic and foreign microtargeted propaganda.

## A. Microtargeted propaganda can be more effective

Take the following hypothetical about traditional propaganda. Suppose that a foreign actor wanted to interfere in a national election and publish an ad in a major newspaper to attack a political candidate on, say, his past vulgar behaviour. There are various factors that limit the effectiveness of such propaganda. First, the ad reaches many people who are not susceptible to the message, because they don't care about politics. Second, the ad reaches people who are not susceptible to the message because they don't care about the vulgar behaviour or don't find the behaviour vulgar. Third, the ad reaches people who do care about the behaviour, but could be influenced more with a different message. Fourth, the ad reaches people who do care a lot about the hypothetical propagandised behaviour and change their opinions accordingly.[41]

Given all these limits, microtargeting can be a more effective way of identifying the 'right' person for propaganda. Microtargeting is based on the premise that the collection of extensive data about the individual through their online activity enables the targeting party to better predict their future activity (from buying patterns to voting preferences) than by using less detailed information such as postcodes, subscriptions or property types used previously by political parties. Actors can then seek to change the behaviour only of population segments especially likely to engage in a particular activity (buying a specific item or voting for a specific party). Microtargeting is therefore widely thought to be more accurate than other forms of differentiation or segmentation between citizens or customers. This assumption underpins political microtargeting and the online advertising industry more broadly.[42]

---

40. D. Flynn et al., 'The Nature and Origins of Misperceptions: Understanding False and Unsupported Beliefs About Politics', 38 *Political Psychology* (2017), p. 127; S. Bradshaw and P.N. Howard, 'The Global Organization of Social Media Disinformation Campaigns', 7 (Special) *Journal for Internal Affairs* (2018), p. 23.
41. A fifth factor that might limit the effectiveness of such propaganda is that many media might refuse publishing such an ad. This relates to the norms and regulation of mass media and social media companies considered in the following sections.
42. Some commentators doubt the effectiveness of targeted advertising. See e.g. G. Edelman, 'Can Killing Cookies Save Journalism?' *Wired*, 5 August 2020, www.wired.com/story/can-killing-cookies-save-journalism/.

Due to this accuracy in audience segmentation and behaviour prediction, microtargeting techniques could enable foreign actors to spread propaganda more effectively by introducing a *customized propaganda narrative*: a message tailored to specific characteristics of subgroups within society. For example, foreign actors could send a person who is concerned about immigration a series of targeted 'news articles' describing the approaching arrival of tens of thousands of immigrants.[43] Because the messages are personally relevant, people are more likely to be influenced. Indeed, Endres found that microtargeting people with political messages on specific issues increased support for the candidate sending the message and decreased support for the opponent.[44] Compared to traditional propaganda, then, microtargeted propaganda is likely to be more effective.

## B. Microtargeted propaganda can be more efficient

The second element is efficiency in delivering the 'right message'; the ease of switching between customised propaganda narratives, in addition to the ability to craft them in the first place. While political parties have long sent different messages to different constituencies based on relatively coarse, publicly available data, microtargeting enables these to be sent simultaneously and at a fine-grained level. In addition to the anti-immigration message above, foreign actors could alternatively send targeted propaganda about a specific candidate based on his or her stance on immigration (this depends on the goals of the propaganda actor: to attack a political candidate, or to bolster support for a specific political party).

Microtargeting also enables propaganda to be honed gradually using A/B testing (comparing the impact of multiple similar adverts) before serving these adverts to a larger population. This ease of switching and replacing messages (for example, the candidate-focused or issue-focused ads above) is qualitatively different to previous forms of political advertising, as it relies on near instantaneous judgement of the 'right' person through online ad auctions and digital, often automated, changes to the advertisement itself. In comparison with traditional propaganda, microtargeted propaganda is thus more efficient.

This concept of efficiency applies only to the ease with which customized propaganda narratives can be changed, duplicated or replaced. It does not seek to address whether microtargeted propaganda is efficient in the broader but equally important sense of return on investment: for a given euro or dollar, how much influence do you purchase with microtargeted propaganda compared to traditional means? The shift in the advertising industry towards microtargeted online advertising suggests that this broader concept of efficiency is also met by microtargeting. However, it would require difficult-to-access data on internal spending over time and across different forms of attempted influence to assess whether microtargeted political propaganda is efficient in this larger sense, especially for foreign actors.

## C. Microtargeted propaganda is often opaque

The opacity of microtargeting further complicates things, as opacity allows foreign actors to pursue anti-democratic goals in the shadows. Together with microtargeted propaganda's effectiveness and efficiency, this opacity makes it difficult to monitor, factcheck and counter microtargeted propaganda.[45] Traditional propaganda, especially by foreign actors, often sought to hide its origin.[46] For microtargeted propaganda, this opacity can be split into two distinct types.

---

43. 'Concern' about immigration is defined by predicted behaviour, i.e. the person whose voting decision is determined by immigration.
44. K. Endres, 'Targeted Issue Messages and Voting Behavior', 48 *American Politics Research* (2019), p. 317.

First, there is what academic research on cybersecurity calls 'the attribution problem': identifying the ultimate source of a specific action or campaign.[47] As this research strand demonstrates, the attribution problem is a long-standing aspect of digital or offline covert operations. Attribution relies on inferred motives, technical indicators and wider 'techniques, practices and procedures' (TTPs) to first group different actions as all originating from the same actor, and then uncover the identity of that actor. This technique is widely applied in disinformation research, as specific pieces of disinformation are traced back to false identities on social networks, then these identities are grouped together in what Facebook calls 'coordinated inauthentic behavior', and finally both 'within-network' and wider indicators, such as financial transfers, are used to identify the original actor.

The complexity of this technique is exemplified by the tracing of apparent Ghanaian NGO activity to companies connected to the same Russian individual associated with the Internet Research Agency that interfered with the 2016 US Presidential election.[48] Microtargeted propaganda inherits all the difficulties of attribution in disinformation and cybersecurity more broadly.

Second, there is another form of opacity, more basic than the attribution problem: it is often not clear exactly what are instances of microtargeted propaganda. While traditional means of propaganda are relatively stable, and institutions exist to document, record and store newspapers, radio and television broadcasts (or even more ephemeral forms of propaganda like campaign leaflets), this activity of recording is more difficult for microtargeted propaganda. While researchers or observers may investigate targeted propaganda in pre-social media by analysing newspapers, this is much more challenging on social networks; only the targeted groups see the messages and they are not served to anyone else.

One possible solution is for researchers to create fake personas matching the characteristics of targeted groups, similar to the technique of creating 'honeypots' in cybersecurity, but this would contravene terms and conditions of the social network sites. Another more promising solution is publicly available advertisement (ad) libraries, which are designed to make microtargeting less opaque. However, these libraries have important shortcomings.[49] Ad libraries are not always accurate and give limited information about the targeting criteria that were used.[50] Consequently, it is difficult to monitor who is exposed to what message with what effect, in addition to the attribution problem above. The ad libraries' shortcomings are problematic, because foreign actors looking to spread propaganda are likely to profit from the opacity of microtargeting to avoid detection, journalistic scrutiny and countermeasures. Overall, the opacity of microtargeted propaganda makes it an attractive choice, especially for foreign actors seeking to avoid attribution.

---

45. Hameleers and Van der Meer show that while factchecks do not completely mitigate disinformation effects, they do have some success in limiting effects of disinformation: M. Hameleers and T. van der Meer, 'Misinformation and Polarization in a High-Choice Media Environment: How Effective Are Political Factcheckers?' 47 *Communication Research* (2020), p. 227.
46. For many examples, see T. Rid, *Active Measures: The Secret History of Disinformation and Political Warfare* (Profile Books, 2020).
47. See e.g. F.J. Egloff, 'Public Attribution of Cyber Intrusions', 6 *Journal of Cybersecurity* (2020), p. 1; T. Rid and B. Buchanan, 'Attributing Cyber Attacks', 38 *Journal of Strategic Studies* (2015), p. 4; and M. Schulzke, 'The Politics of Attributing Blame for Cyberattacks and the Costs of Uncertainty', 16 *Perspectives on Politics* (2018), p. 954.
48. C. Ward et al., 'How Russian Meddling Is Back before 2020 Vote', *CNN*, 13 March 2020, www.cnn.com/2020/03/12/world/russia-ghana-troll-farms-2020-ward/index.html.
49. P. Leerssen et al., 'Platform Ad Archives: Promises and Pitfalls', 8 *Internet Policy Review* (2019), p. 1; and R. Tromble et al., *Transparency in Digital Political Advertisements during the 2019 European Parliament Elections* (Netherlands Helsinki Committee and the European Partnership for Democracy, 2019).
50. P. Leerssen et al., 8 *Internet Policy Review* (2019), p. 1.

## D. Microtargeted propaganda can be combined with other technologies

Finally, microtargeted propaganda can be combined with other relevant technological advances, such as deepfakes. Deepfakes are manipulated videos that can make it seem as if a person says or does something, while in reality the depicted person never did so. For instance, during the 2019 election campaign in the UK, Boris Johnson was seen and heard endorsing his competitor Jeremy Corbyn. In reality, as the deepfake video later showed, this never happened.[51] Deepfakes are based on real images. Making a deepfake requires some work, but once the required neural network is trained it is relatively easy to produce (slightly) different deepfakes.

This section focuses on deepfakes because they are understood to be a potentially powerful mode of propaganda. Doctored images and news articles are also relevant. However, unlike text- or image-based propaganda, deepfakes deceive the eyes *and* the ears. Deepfakes have been shown to be effective in decreasing political attitudes toward a politician negatively depicted in a deepfake.[52] And deepfakes have been shown to poison the public debate by sowing confusion about what is real, and consequently reducing trust in news on social media.[53] The less people trust news media, the more likely they are to fall for propaganda which, in turn, can affect their voting behaviour.[54] Microtargeting can further amplify negative effects of deepfakes, making microtargeting a powerful propaganda tool.[55] For example, a French dairy farmer would likely react more strongly to a deepfake in which EU Commission President Ursula von der Leyen ostensibly unfolds a plan to cut down farmer subsidies, than to a deepfake in which she ostensibly criticizes the techniques used by Dutch pulse fishermen. For Dutch fishermen, the opposite is likely true. Both elements of microtargeting considered above amplify the effects of deepfakes: the greater ability to identify suitable targets for propaganda (effectiveness), and the ability to serve different deepfakes, possibly with only subtle changes, to different population segments simultaneously (efficiency), with swift refinements to hone the deepfake if it appears unsuccessful. Consequently, sending a deepfake may already have negative and damaging effects, but microtargeting can amplify these effects by enabling the spread of tailored deepfakes.

Overall, microtargeted propaganda has several novel characteristics: greater effectiveness, greater efficiency, more opacity and the potential for combination with other new forms of technological influence. These are extended or exacerbated features of propaganda delivered by traditional means rather than entirely new.

51. S. Cole, 'Deepfake of Boris Johnson Wants to Warn You about Deepfakes', *Vice*, 12 November 2019, www.vice.com/en_us/article/8xwjkp/deepfake-of-boris-johnson-wants-to-warn-you-about-deepfakes.
52. T. Dobber et al., 'Do (Microtargeted) Deepfakes Have Real Effects on Political Attitudes?', *International Journal of Press/Politics* (2020), p. 1.
53. C. Vaccari and A. Chadwick, 'Deepfakes and Disinformation: Exploring the Impact of Synthetic Political Video on Deception, Uncertainty, and Trust in News', *Social Media and Society* (2020), p. 1.
54. See F. Zimmermann and M. Kohring, 'Mistrust, Disinforming News, and Vote Choice: A Panel Survey on the Origins and Consequences of Believing Disinformation in the 2017 German Parliamentary Election', 37 *Political Communication* (2020), p. 215.
55. T. Dobber et al., *International Journal of Press/Politics* (2020), p. 1.

## 4. What can lawmakers do to mitigate the risks?

This section highlights possible policy responses to the risks associated with microtargeted propaganda by foreign actors. First, we discuss how privacy and data protection law could help to alleviate the threat of microtargeted propaganda. Next, we explore additional rules that policymakers could adopt.

### A. Data protection law

Enforcement of data protection law, such as the General Data Protection Regulation (GDPR) in Europe, could help to mitigate the threat of microtargeted foreign propaganda. The GDPR imposes obligations on organisations that use personal data ('data controllers'),[56] and gives rights to people whose personal data are used ('data subjects').[57] Independent Data Protection Authorities oversee compliance.[58] The GDPR applies as soon as personal data ('information relating to an identified or identifiable natural person') are processed. The GDPR defines processing broadly; almost everything that can be done with personal data falls within the processing definition.[59]

Roughly summarised, some of the main principles of the GDPR are as follows. Personal data may only be used fairly and transparently ('lawfulness, fairness and transparency'). Personal data that are collected for one purpose, may not be used for unrelated other purposes ('purpose limitation'). Data controllers must not collect or use more personal data than is necessary for the processing purpose ('data minimisation'), and may not be retained for an unreasonably long period ('storage limitation'). Personal data must be reasonably accurate and up to date ('accuracy'), and must be kept securely ('integrity and confidentiality'). Data controllers are responsible for compliance ('accountability').[60]

The GDPR is long, detailed and nuanced, and a full discussion of the interplay between the GDPR and microtargeted propaganda falls outside the scope of this article.[61] We highlight a few examples of GDPR rules that could help to protect people and democratic societies against negative effects of microtargeted propaganda by foreign actors.

First, the GDPR makes microtargeting in the EU more difficult and more expensive than in, for instance, the US, which does not have a GDPR-like statute. For example, in the US, many files on individuals can be bought from data brokers. Partly because of stricter data protection rules, it is more difficult in Europe to buy personal data about individuals.[62] And in most EU countries it is

---

56. Regulation (EU) 2016/679 of the European Parliament and of the Council of 27 April 2016 on the protection of natural persons with regard to the processing of personal data and on the free movement of such data, and repealing Directive 95/46/EC ('GDPR'), [2016] OJ L 119/1, Article 4(7).
57. GDPR, Article 4(1).
58. GDPR, Chapter VI.
59. GDPR, Article 4(2).
60. GDPR, Article 5.
61. See for more details: European Commission, 'Commission guidance on the application of Union data protection law in the electoral context', A contribution from the European Commission to the Leaders' meeting in Salzburg on 19–20 September 2018, COM/2018/638 final. See also C. Bennett and S. Oduro-Marfo, 'Privacy, Voter Surveillance and Democratic Engagement: Challenges for Data Protection Authorities', International Conference of Data Protection and Privacy Commissioners (October 2019), http://dx.doi.org/10.2139/ssrn.3517889.
62. Bennett suggests that the fact that online political microtargeting happens so much in the US can be partly explained by the absence of a general data protection law in the US; C. Bennett, 'Voter Databases, Micro-targeting, and Data Protection Law: Can Political Parties Campaign in Europe as they do in North America?', (2016) 6 *International Data Privacy Law* (2016), p. 261.

impossible to buy voter files about people.[63] In the US, party affiliations of people are often public.[64]

Propagandists can, however, target advertising without collecting their own data, for instance by advertising via Facebook. It seems questionable whether Facebook fully complies with the GDPR. Facebook faces regular legal challenges in Europe because of data protection law, but enforcement of the GDPR against Facebook could be stronger.

Second, the GDPR has rules that require transparency about personal data usage. The GDPR requires the controller (such as the advertiser and the social media company[65]) to disclose information such as its identity and the processing purpose.[66] Such information must be given in 'a concise, transparent, intelligible and easily accessible form'.[67] Hence, under the GDPR, people who receive targeted propaganda should legally be provided with the identity of the propagandist and the goals of the personal data use by the propagandist.

Third, for targeted advertising, the GDPR typically requires the data subject's prior consent. The GDPR does not always require prior consent for the use of personal data; there are six possible legal bases for personal data processing; and consent is one of them. However, scholars and Data Protection Authorities generally agree that the data subject's prior consent is required for targeted advertising.[68] Especially when 'special categories of data', such as data about people's 'political opinions' are used, the GDPR generally requires prior consent of the data subject.[69] In addition, the ePrivacy Directive requires the individual's prior informed consent for the use of tracking cookies and similar tracking techniques – essential techniques for many types of targeted advertising.[70]

In sum, if properly enforced, the GDPR could help to protect people and democracies against negative effects of microtargeted propaganda. As the European Commission notes,

> Protecting personal data is (…) instrumental in preventing the manipulation of citizens' choices, in particular via the micro-targeting of voters based on the unlawful processing of personal data, avoiding interference in democratic processes and preserving the open debate, the fairness and the transparency that are essential in a democracy.[71]

---

63. Ibid.
64. E. Hersh, *Hacking the Electorate: How Campaigns Perceive Voters* (Cambridge University Press, 2015).
65. Whether an advertiser and Facebook must be seen as (joint) controllers requires a nuanced and facts-specific assessment, which falls outside the scope of this article. As the European Commission notes, '[i]n the electoral context, a number of actors can be data controllers: political parties, individual candidates and foundations are, in most instances, data controllers; platforms and data analytics companies can be (joint) controllers or processors for a given processing depending on the degree of control they have over the processing concerned' (European Commission, 'Commission guidance on the application of Union data protection law in the electoral context', A contribution from the European Commission to the Leaders' meeting in Salzburg on 19–20 September 2018, COM/2018/638 final, p. 4). See on joint controllership also: R. Mahieu and J. van Hoboken, 'Fashion-ID: Introducing a Phase-Oriented Approach to Data Protection?', *European Law Blog* (2019), https://europeanlawblog.eu/2019/09/30/fashion-id-introducing-a-phase-oriented-approach-to-data-protection/
66. GDPR, Article 13(1)(a).
67. GDPR, Article 12(1)(a).
68. F.J. Zuiderveen Borgesius, 'Personal Data Processing for Behavioural Targeting: Which Legal Basis?', 5 *International Data Privacy Law* (2015), p. 163.
69. GDPR, Article 9(1).
70. Article 5(3) of the ePrivacy Directive, Directive 2002/58/EC of the European Parliament and of the Council of 12 July 2002 concerning the processing of personal data and the protection of privacy in the electronic communications sector (Directive on privacy and electronic communications), as amended by Directive 2006/24/EC [the Data Retention Directive], and Directive 2009/136/EC [the Citizen's Rights Directive].

We do not suggest that the GDPR could solve the problems related to microtargeted propaganda. It is well-documented that legal requirements of transparency and consent have only limited effect. Most people don't read privacy notices, and click 'I agree' without thinking.[72] Moreover, there is a compliance and an enforcement deficit.

Nevertheless, it seems plausible that data protection law in Europe hinders microtargeting. One political campaign leader in the Netherlands said: 'you never want to abuse someone's personal data. So yes, regulations sometimes cause us to hit the brake and that's a good thing'.[73] In sum, if properly enforced, the GDPR could be a useful tool to alleviate the risks of microtargeted propaganda. For countries without GDPR-like rules on the books, adopting such rules would be an essential step towards a regulatory environment that can protect their society against microtargeted propaganda. Next, we discuss options for policymakers in the EU that consider additional rules.

## B. Options for additional rules

*1. Banning or limiting propaganda from foreign actors.* The most draconian option would be to ban microtargeted propaganda outright, including by both domestic and foreign actors. Such a ban could be framed along the lines of the current wave of false information laws that are sweeping across the world. For example, under Singapore's Protection from Online Falsehoods and Manipulation Act 2019, it is an offence to communicate a false statement that is 'likely' to influence the outcome of an election or a referendum, or 'incite feelings of enmity, hatred or ill-will' between different groups of people, or 'diminish public confidence' in the performance of the Singapore government, or government agency.[74]

Such a broad provision would likely capture many of the microtargeted propaganda messages by foreign actors seeking to influence voters in elections, incite feelings of enmity between different social groups, or undermine government policies. Further, it is also specific offence to use a bot, 'whether in or outside Singapore', for spreading such false statements.[75] A bot is defined under the 2019 Act as a computer program made or altered for the purpose of running automated tasks.[76] Such a broadly worded provision would not only apply to foreign actors, but also to domestic actors engaging in similar propaganda, whether at the behest of, or separate to, foreign actors. As such, this would address a problem that may arise where legislation only targets foreign actors, as is could lead to the concealing of the sources, e.g., using domestic proxies for dissemination.

---

71. European Commission, Communication from the Commission to the European Parliament and the Council, Data protection as a pillar of citizens' empowerment and the EU's approach to the digital transition – two years of application of the General Data Protection Regulation, COM(2020) 264 final, 26 June 2020, p. 16.
72. A. Acquisti et al., 'Privacy and Human Behavior in the age of Information', 347 *Science* (2015), p. 509; S.C. Boerman et al., 'Online Behavioral Advertising: A Literature Review and Research Agenda', 46(3) *Journal of Advertising* (2017), p. 363; and P. Grassl et al., 'Dark and Bright Patterns in Cookie Consent Requests', 3(1) *Journal of Digital Social Research* (2021), https://doi.org/10.33621/jdsr.v3i1.54.
73. T. Dobber et al., 'Two Crates of Beer and 40 Pizzas: the Adoption of Innovative Political Behavioural Targeting Techniques', 6 *Internet Policy Review* (2017), p. 1.
74. Protection from Online Falsehoods and Manipulation Act 2019, section 7.
75. Ibid., section 8.
76. Ibid., section 2(1).

However, such a policy option raises acute freedom of expression issues, which are detailed below, in section 5.

A more targeted form of regulation could be framed along the lines of Canada's Elections Modernization Act 2018, which sought to prohibit online political advertising from certain foreign actors. Under the 2018 Act, it is prohibited to sell any advertising space to a foreign political party or a foreign government or an agent or mandatary of a foreign government 'for the purpose of enabling that person or entity to transmit an election advertising message or to cause an election advertising message to be transmitted'.[77]

This type of regulation would raise problems in the context of EU member states, in that it would prohibit a Dutch political party from buying online ads in Belgium or France, to support a similarly-aligned party. In Europe, national political parties are part of pan-European political parties that contest the European Parliament election, such as the right-leaning European People's Party and the left-leaning Party of European Socialists.

A related approach may be to prohibit foreign political microtargeting from outside the EU, but allow political microtargeting between EU member states. Thus, Russian organisations would be prohibited from buying microtargeting advertising targeted individuals in the EU member states, but an organisation based in one EU member states (for example a Dutch NGO) would be permitted to buy microtargeted advertising targeting individuals in Germany. Regulation of this sort is being implemented in the Netherlands in the area of campaign finance regulation. Under an amended Political Parties Financing Act, financial donations from outside the EU to Dutch political parties are banned (unless from a Dutch citizen abroad), while donations from within the EU are permitted, but require full transparency.[78]

However, a complicating factor for such a regulation applying to political microtargeting is that all EU member states are also members of the Council of Europe, a larger 47-country international organisation, which includes non-EU countries such as Russia, Ukraine, and the United Kingdom. This means that all EU member states are subject to the European Convention on Human Rights, which guarantees freedom of expression across borders between its member states.[79]

Notably, one method that has been adopted which avoids *legislating* to impose prohibitions on microtargeting propaganda from foreign actors in to instead place such prohibitions with *voluntary* codes of conducts between governments and online platforms. For example, in March 2021, in the run-up to the Dutch parliamentary elections, a voluntary Code of Conduct on transparency in online political advertisements was agreed between the Dutch Ministry of the Interior and Kingdom Relations, a number of online platforms, Dutch political parties and the International Institute for Democracy and Electoral Assistance (an intergovernmental organisation to support and strengthen democratic institutions).[80] Notably, the Dutch political parties committed to refuse direct purchases of political advertisements by foreign actors in support of the political party (with and without attributing them to the party); and to refrain from receiving foreign funding to pay for online political advertisements, other than from party members living abroad.[81] Further, the online platforms agreed to '[b]an cross-border political advertisements from outside the European Union'.[82] However, the major drawback with these code of conduct is that they lack transparency in terms actual implementation, and also delegating regulation to platforms.[83]

---

77. Elections Modernization Act 2018, section 282.4(5).
78. 'Ministerraad stemt in met wijziging van de Wet financiering politieke partijen' ['Cabinet approves amendment of the Financing of Political Parties Act'], 7 February 2020, www.rijksoverheid.nl/actueel/nieuws/2020/02/07/ministerraad-stemt-in-met-wijziging-van-de-wet-financiering-politieke-partijen.
79. Convention for the Protection of Human Rights and Fundamental Freedoms, 4 November 1950, ETS No.005.
80. Ministry of the Interior and Kingdom Relations, Dutch Code of Conduct Transparency Online Political Advertisements (International Institute for Democracy and Electoral Assistance, 2021), www.idea.int/sites/default/files/news/news-pdfs/Dutch-Code-of-Conduct-transparency-online-political-advertisements-EN.pdf.

*2. Improving transparency.* An additional approach would seek to bring transparency to microtargeted propaganda, allowing journalists, regulators, and others to see who is paying for ads from abroad. Regulation could, for instance, impose an obligation on those purchasing ads to include a disclaimer on their identity, or an obligation on online platforms to verify the identity of those purchasing ads.[84]

France has chosen the latter approach. The French 2018 Law on Manipulation of Information obliges online platforms to provide users with 'fair, clear and transparent' information on the identity of the person or group which has paid for promoted content relating to a 'debate of general interest'.[85] This obligation only applies during election periods in France, and is designed to bring transparency to all paid-for content that concerns a matter of general interest, and covers a great deal more content than regulations targeting merely ads concerning a political candidate. Further, in the United States, a number of states have enacted legislation designed to introduce transparency for political advertising online.[86] For example, California enacted the Social Media Disclosure Act, which requires political advertising on online platforms to include a disclosure on who paid for the advertisement; and requires platforms to keep a publicly available database of the political ads.[87] And under federal law in the US, certain paid political advertising is required to include a 'disclaimer' identifying the person who paid for a communication and whether the communication was authorised by a political candidate.[88] Indeed, the US Supreme Court has held that held such transparency rules do not violate the right to freedom of speech under the First Amendment.[89]

## 5. Complications when mitigating the risks

Lawmakers in Europe run into several difficulties when they try to mitigate the risks of microtargeted propaganda.

---

81. Ibid., section 3.2.
82. Ibid., para. 18.
83. See Report of the Special Rapporteur on the promotion and protection of the right to freedom of opinion and expression, A/HRC/38/35, 6 April 2018, para. 19–20 (noting that non-binding agreements with platforms 'have limited transparency' and 'companies perform public functions without the oversight of courts and other accountability mechanisms').
84. The European Commission has indicated it will present a legislative proposal on the transparency of sponsored political content in 2021. See European Commission, Communication on the European Democracy Action Plan, COM/2020/790 final, p. 4.
85. Loi n° 2018-1202 du 22 décembre 2018 relative à la lutte contre la manipulation de l'information, article 1.
86. See Joris van Hoboken et al., The Legal Framework on the Dissemination of Disinformation through Internet Services and the Regulation of Political Advertising (A report for the Ministry of the Interior and Kingdom Relations, 2019), p. 155.
87. An act to amend Sections 84504.3, 84504.4, and 84510 of, and to add Sections 84503.5 and 84504.6 to, the Government Code, relating to the Political Reform Act of 1974.
88. See 11 CFR § 110.11 – Communications; advertising; disclaimers (52 U.S.C. 30120).
89. See *Citizens United v. Federal Election Commission*, 558 U.S. 310 (2010), 366 (holding that '[d]isclaimer and disclosure requirements may burden the ability to speak, but they "impose no ceiling on campaign related activities," and "do not prevent anyone from speaking"').

### A. The right to freedom of expression

When regulating propaganda or advertising, lawmakers must consider freedom of expression. Under international human rights law, freedom of expression is guaranteed 'regardless of frontiers', and freedom of expression standards limit any restrictions not only within a jurisdiction, 'but also those which affect media outlets and other communications systems operating from outside of the jurisdiction of a State as well as those reaching populations in States other than the State of origin'.[90]

Under the European Convention on Human Rights, Article 10 guarantees the right to 'receive and impart information and ideas without interference by public authority and regardless of frontiers'. The European Court of Human Rights has held that the right to freedom to receive information prohibits a government from restricting a person from receiving information that others wish to impart to him or her. Thus, imposing a restriction on information from abroad is an interference with freedom of expression. To illustrate: in *Association Ekin v. France*, the European Court of Human Rights considered a French law which permitted the banning of publications of 'foreign origin'.[91] The Court found that a ban imposed on a book's publication under the provision violated Article 10, and held that '[s]uch legislation appears to be in direct conflict with the actual wording of paragraph 1 of Article 10 of the Convention, which provides that the rights set forth in that Article are secured regardless of frontiers'.[92] Indeed, this fundamental principle of freedom of expression law has been applied by the Court to instances of blocking 'illegal' online content from 'foreign-based' platforms, such as in the Court's unanimous judgments in *Ahmet Yıldırım v. Turkey*,[93] (concerning Google Sites) and *Cengiz and Others v. Turkey* (concerning YouTube).[94]

Political advertising is a form of political communication, and thus, is an exercise of the right to freedom of expression, which is guaranteed by both Article 11 of the EU Charter of Fundamental Rights, and Article 10 of the European Convention on Human Rights.[95] Because political advertising is considered political expression, it enjoys a 'privileged position' under Article 10.[96] Because of that privileged position, the European Court of Human Rights applies its highest standard of scrutiny – strict scrutiny – to any restriction on political speech. Because there is 'little scope' for restrictions on political speech, any restriction must be 'narrowly interpreted', and its necessity 'convincingly established' by the government.[97]

### B. Comparison with rules on TV broadcasting

In the context of TV broadcasting, there is some experience with limiting expression that originates from foreign actors. Under international human rights law, jamming signals from a broadcaster based in another jurisdiction, or withdrawing rebroadcasting rights in relation to that broadcaster's programmes, is legitimate 'only where the content disseminated by that broadcaster has been held by a court of law or another independent, authoritative and impartial oversight body to be in serious and persistent breach of a legitimate restriction on content'.[98]

---

90. Joint Declaration, 2017, section 1(d).
91. ECtHR, *Association Ekin v. France*, Judgment of 17 July 2001, Application No. 39288/98).
92. Ibid., para. 62.
93. ECtHR, *Ahmet Yıldırım v. Turkey*, Judgment of 18 December 2012, Application No. 3111/10, para. 67.
94. ECtHR, *Cengiz and Others v. Turkey*, Judgment of 1 December 2015, Application Nos. 48226/10 and 14027/11, para. 65.
95. See also T. Dobber et al., 'The Regulation of Online Political Microtargeting in Europe', 8 *Internet Policy Review* (2019), p. 7.
96. ECtHR, *TV Vest As & Rogaland Pensjonistparti v. Norway*, Judgment of 11 December 2008, Application No. 21132/05, para. 66.
97. ECtHR, *Vitrenko and Others v. Ukraine*, Judgment of 16 December 2008, Application No. 23510/02), para. 1.

EU law is also informative for the discussion of political microtargeting by foreign actors, in particular broadcasting and audiovisual media law, which is regulated under the EU's Audiovisual Media Services Directive.[99] That AVMS Directive says that EU member states must not generally 'restrict retransmission' of audiovisual media services from other EU member states.[100] The Directive provides a mechanism for a derogation from this rule in certain limited circumstances. Thus, a member state may restrict broadcasts from another member state where a media service provider 'manifestly, seriously and gravely infringes' the prohibition on 'incitement to violence or hatred', or 'prejudices or presents a serious and grave risk of prejudice to public security, including the safeguarding of national security and defence'.[101] However, there is a detailed mechanism, with many hurdles, under the AVMS Directive before an EU member state can restrict a transmission from abroad.[102]

## C. European Union law

When regulating political microtargeting, policymakers may want to aim regulation at platforms providers, such as Facebook, Twitter, and YouTube. However, EU law contains a specific legal regime for such companies, when these companies fall within the legal category of 'hosts'.

The e-Commerce Directive has a special exemption from liability for hosts, in short internet companies that offer a service that consists of the storing, or hosting, information for others.[103] Roughly summarised, Article 14 of the Directive requires EU member states to ensure that the hosting provider is not liable for the information stored at the request of a user, if two conditions are met: (a) the provider does not have actual knowledge of illegal information; and (b) if the provider realizes that it stores illegal content (for instance after receiving a credible take down request) it quickly takes down that content. Hence, if a provider meets those two conditions, it can rely on the safe harbour offered by article 14, protecting the provider from liability for illegal content on its platform.

Article 15 lays down a prohibition of general monitoring obligations for hosts: 'Member States shall not impose a general obligation on providers (…) to monitor the information which they transmit or store, nor a general obligation actively to seek facts or circumstances indicating illegal activity'. Hence, hosting providers do not have a general obligation to monitor their own platforms to look for, or take down illegal content.

---

98. Joint Declaration, 2017, section 1(h).
99. Directive (EU) 2018/1808 of the European Parliament and of the Council of 14 November 2018 amending Directive 2010/13/EU on the coordination of certain provisions laid down by law, regulation or administrative action in Member States concerning the provision of audiovisual media services (Audiovisual Media Services Directive) in view of changing market realities, [2018] OJ L 303/69 ('AVMSD').
100. AVMSD, article 3(1).
101. Ibid., article 3(2).
102. See ibid., article 3(1)–(7).
103. Directive 2000/31/EC of the European Parliament and of the Council of 8 June 2000 on certain legal aspects of information society services, in particular electronic commerce, in the Internal Market ('Directive on electronic commerce'), [2000] OJ L 178/1.

The Directive was adopted almost 20 years ago, when the EU lawmaker mainly thought of more traditional hosting services like hosting websites. If hosts were fully liable for hosting illegal content, thought the EU lawmaker, fewer companies would offer hosting services, because of the legal risks involved. The idea was that a lack of hosting services would be bad for business and freedom of expression.[104]

Nowadays, platforms such as Facebook and YouTube can also rely on the safe harbour for hosts, as the companies store information (such as posts, pictures, and videos) for users. Because of the hosting safe harbour, platforms do not have full legal responsibility for the content that their users put on their platform. And providers do not have a legal obligation to monitor their platforms for illegal content.

In 2018, a number of platform providers, including Facebook, Google and Twitter, signed the EU Code of Practice on Disinformation (2018), a self-regulation instrument. The signatories promised to slow down the spread of online disinformation and fake news.[105]

There is a trend towards giving more legal responsibilities to hosting and platform providers. Already, in some sectors, EU policymakers aim to make platform providers more responsible for content. For instance, the EU Terrorist Content Online Regulation contains rules on duties of care to be applied by hosting providers to prevent the dissemination of terrorist content.[106] In theory, the EU could adopt rules that impose more responsibilities on platforms and hosts, for instance to make them stop the circulation of certain types of disinformation or propaganda.

However, there are good arguments for not making platform providers fully responsible for the information (including microtargeted propaganda) that they host. First, it is questionable whether platforms would ever be able to ensure full compliance, seeing the amount of information that users upload every second. Second, and more importantly, if the law made platforms fully liable for information they host, they might decide to ban all political discussion. Such an outcome would be bad for freedom of expression, and would be a bad outcome for our democratic society. Nevertheless, it may be possible to strike a balance. The law could make platform providers more responsible than now, while not imposing full responsibility for what happens on their platforms.

Indeed, in December 2020, the European Commission published a proposal for a new Digital Services Act (DSA), which is designed to place a raft on new obligations on online platforms in in order to ensure a 'safe, predictable and trusted online environment', and so that fundamental rights are 'effectively protected' online.[107] Notably, under Article 26 DSA, certain large online platforms will be required to carry out risk assessments of 'significant systemic risks' stemming from the functioning of their services, including the 'intentional manipulation of their service, including by means of inauthentic use or automated exploitation of the service, with an actual or foreseeable negative effect on the protection of public health, minors, civic discourse, or actual or foreseeable effects related to electoral processes and public security'.[108]

---

104. See J. van Hoboken, *Search Engine Freedom: On the Implications of the Right to Freedom of Expression for the Legal Governance of Web Search Engines* (Kluwer Law International BV, 2012), p. 138.
105. EU Code of Practice on Disinformation (2018) https://ec.europa.eu/newsroom/dae/document.cfm?doc_id=54454. See P.H. Chase, 'The EU Code of Practice on Disinformation: The Difficulty of Regulating a Nebulous Problem', *Working Paper of the Transatlantic Working Group on Content Moderation Online and Freedom of Expression* (2019).
106. Regulation (EU) 2021/784 of the European Parliament and of the Council of 29 April 2021 on addressing the dissemination of terrorist content online [2021] OJ L 172/79, Article 1(1)(a).
107. Proposal for a Regulation on a Single Market For Digital Services (Digital Services Act) and amending Directive 2000/31/EC, COM(2020) 825 final, 15 December 2020.
108. Proposal for a DSA, Article 26.

Notably, Recital 57 DSA gives further examples of intentional manipulation of a platform's service, such as the 'creation of fake accounts, the use of bots, and other automated or partially automated behaviours, which may lead to the rapid and widespread dissemination of information that is illegal content or incompatible with an online platform's terms and conditions'.[109] Crucially, under Article 27 DSA, certain platforms will be required to put in place 'effective mitigation measures' tailored to these systemic risks.[110] These measures can include platforms 'adapting content moderation or recommender systems', adapting 'terms and conditions', targeting measures to limit the display of advertisements and 'reinforcing' internal processes or supervision of their activities, in particular as regards detection of systemic risk.[111] While the DSA makes no mention of propaganda nor foreign interference, it may go some way in mitigating techniques used for the distribution of microtargeted propaganda, such as through the use of fake accounts.

Further, under Article 30 DSA, certain online platforms will be required to establish publicly-available depositories containing all online advertisements displayed on their platforms.[112] Recital 63 DSA explains the purpose of these advertisement depositories is to facilitate supervision and research in relation to risk associated with online advertisements, including 'manipulative techniques and disinformation with a real and foreseeable negative impact on public health, public security, civil discourse, political participation and equality'.[113]

In addition, another recent landmark legal proposal published by the European Commission in 2021, setting down rules on the use of artificial intelligence systems (Artificial Intelligence Act),[114] is also important to briefly mention. Notably, the AI Act would introduce certain transparency requirements for certain AI systems, including systems used for deep fakes. In this regard, Article 52 provides that users of AI systems that generate deep fakes 'shall disclose that the content has been artificially generated or manipulated'.[115] However, there are exceptions to this obligation where use of the deep fake is 'necessary for the exercise of the right to freedom of expression and the right to freedom of the arts and sciences'.[116]

### D. EU competences

Another complication concerns the competences of different law-making bodies. The EU has competence to regulate personal data, and to regulate elections for the European Parliament, but no specific competence to regulate national elections. Hence, national parliaments seem best placed to regulate political microtargeting.

However, EU law does regulate the operation and funding of European political parties and affiliated foundations involved in elections to the European Parliament, which take place every five years.[117] In 2019, the EU enacted an amendment that introduced an obligation on European political parties. Such parties must not 'deliberately influence, or attempt to influence, the outcome of elections to the European Parliament by taking advantage of an infringement' of data protection rules.[118]

---

109. Ibid., Recital 57
110. Ibid., Article 27.
111. Ibid., Article 27(a)–27(b).
112. Ibid., Article 30.
113. Ibid., Recital 63.
114. Proposal for a Regulation laying down harmonised rules on artificial intelligence (Artificial Intelligence Act) and amending certain Union legislative acts, COM(2021) 206 final, 21 April 2021.
115. Proposal for an Artificial Intelligence Act, Article 52(3) ('Users of an AI system that generates or manipulates image, audio or video content that appreciably resembles existing persons, objects, places or other entities or events and would falsely appear to a person to be authentic or truthful ("deep fake"), shall disclose that the content has been artificially generated or manipulated')
116. Ibid., Article 52(3).

## 6. Conclusion

This article explored in what ways microtargeted propaganda by foreign actors is different from more traditional forms of foreign propaganda. New characteristics of microtargeted propaganda, compared to traditional propaganda, can be grouped in three categories.

First, compared to traditional propaganda, microtargeting can be more effective. The sender does not have to spend money and time on reaching people who are not susceptible to its messages. And microtargeting enables the sender to adapt messages to the interests or fears or certain groups in the receiving country. Second, microtargeting can be more efficient in terms of switching or changing messages based on immediate feedback, as well as potentially a broader concept of efficiency based on return on investment. Third, microtargeted propaganda can remain more hidden than traditional propaganda. Like traditional propaganda, the originating actor is often obscure, and the 'attribution problem' is inherited by microtargeted propaganda from longer histories of intelligence and covert action. Furthermore, with mass media propaganda, journalists, the state and others in the receiving country could see the propaganda, and offer a counter-narrative. It is more difficult to monitor microtargeted propaganda. Only a few groups in the receiving country might see the propaganda messages – others in that country may not be aware of those messages.

We also explored what lawmakers in Europe could do to mitigate the risks of microtargeted propaganda. When regulating advertising and propaganda, lawmakers must respect the right to freedom of expression. Nevertheless, lawmakers have some options. First, in specific situations, and only under certain circumstances, lawmakers can limit or even prohibit certain types of propaganda. Second, a slightly lighter measure is limiting or banning certain propaganda types, when they are sent from foreign countries, or when they are sent from foreign countries outside the EU. Again however, such measures would only be legally possible under very limited conditions.

Finally, lawmakers could adopt rules to improve transparency of microtargeted propaganda. For instance, the law could require that political ads (however defined) show who has paid for the ad. The law could also make social media platforms partly responsible for ensuring that such ads are labelled as political ads. Seeing that much is still unknown about microtargeted propaganda, such measures to improve transparency may be the best option in the short term.

### Declaration of Conflicting Interests

---

117. Regulation (EU, Euratom) No 1141/2014 of the European Parliament and of the Council of 22 October 2014 on the statute and funding of European political parties and European political foundations, [2014] OJ L 317/1.
118. Regulation (EU, Euratom) 2019/493 of the European Parliament and of the Council of 25 March 2019 amending Regulation (EU, Euratom) No 1141/2014 as regards a verification procedure related to infringements of rules on the protection of personal data in the context of elections to the European Parliament, [2019] OJ L 851/7, Article 10a.

## Funding

The author(s) received no financial support for the research, authorship and/or publication of this article.

## ORCID iD

Ronan Ó Fathaigh  https://orcid.org/0000-0002-4494-2599